\documentclass[12pt]{article}

\usepackage[]{umlaut,latexsym}
\usepackage[xdvi]{graphicx}

\usepackage{natbib}

\begin{document}

\parskip 2mm


\renewcommand{\refname}{\normalsize \bf \em References}

\title{\bf CRYSTAL STRUCTURES OF Ni\boldmath$_2$MnGa FROM 
DENSITY FUNCTIONAL CALCULATIONS}
%
%
\author{
      A.\ ZAYAK\footnote{Corresponding author:
      Tel.: +49 203 379 1606,
      fax:  +49 203 379 3665,
      email: alexei@thp.uni-duisburg.de}
      , W. A. ADEAGBO
      , P.\ ENTEL, \\
      V.\ D.\ BUCHELNIKOV$^{a,}$
\\*[0.2cm]
      {\small \it Institute of Physics,} \\
      {\small \it University of Duisburg-Essen,
      Duisburg Campus, 47048 Germany}
\\*[0.2cm]
      $^a${\small \it Physics Department,} \\
      {\small \it Chelyabinsk State University,
      454021 Chelyabinsk, Russia}
}
\date{\small \it (Received ...)}
\maketitle

\begin{abstract}
%
The different crystal structures of ferromagnetic Ni$_2$MnGa  
have been calculated using density functional theory (DFT) with special
emphasis on the modulated structures 10M and 14M. These are
important for understanding the stability of Ni$_2$MnGa martensites and
their functionality as shape-memory materials. The modulated structures have
been optimized in the calculations and their properties are discussed in
relation to the structures without modulation. The occurrence of the modulated
structures is related to the soft TA$_2$ phonon mode observed in
Ni$_2$MnGa. The latter is related to the specific nesting behavior of the
Fermi surface in Ni$_2$MnGa. Particular shapes of the modulated structures
are stabilized by the covalent interaction mediated by the \textit{p}-orbitals
of Ga and \textit{d}-orbitals of Ni. The role of this interaction becomes
clear seen when 
considering the phonon dispersion spectrum of Ni$_2$MnGa, where some
characteristic anomalies occur in the coupling of acoustical vibrational modes 
and the optical modes of Ni.
%
\end{abstract}

\noindent {\it Keywords:} Heusler alloys, 
Modulated structures, 
Density functional calculations

\section*{1. INTRODUCTION}

A hundred years have passed since the martensites had been discovered. However,
during these days new martensitic structures are being discovered and special
focus is now on the details of the microscopic nature of this phenomenon. This
is due to the unique dynamical properties of martensitic structures, which are
essentially different from other materials. This property is based on the
mobility of the martensitic 
domains which allows to induce large macroscopic deformations of the sample by
applying an external stress. This deformation does not cost much energy since
the crystal structure remains unchanged; only the domain walls move. In turn,
the high temperature austenitic structure does not show these structural
domains. Therefore, heating up the system allows to get rid of the
deformations supplied by the martensitic state. Whatever was done to the
sample at lower temperatures will be forgotten at higher temperatures. Latter,
one can cool down the sample to the martensitic state and apply new
deformations knowing that everything can be removed as soon as the sample is
heated up to the austenitic state. This is the essence of the shape-memory
behavior which is a completely material dependent property and therefore
requires significant knowledge of the crystals. 

The technological importance of magnetic shape-memory 
materials (MSM) based on Ni$_2$MnGa has been discovered just a few years ago 
\cite{Ullakko-APL96}. The idea here is to use an external magnetic field in
order to induce the motion of martensitic twin boundaries, instead of pressure
as in the non-magnetic martensites. This is possible because of ferromagnetic
order of the martensite structure. In addition, this requires that 
the energy of magnetic anisotropy must be larger than the energy necessary for
twin motion. These conditions are met in Ni$_2$MnGa, but to be more precise, in
its off-stoichiometric compounds. It has been shown that one can achieve
macroscopic deformations of about several percent in the sample by applying
moderate magnetic fields and this can be used to design novel technological
devices like actuators \cite{Aaltio-00}. 

As mentioned above, the MSM effect is a material property. Depending
on the kind of martensite formed at working temperatures the corresponding
effect may be observed. Thus, about 6\% strain was obtained in Ni-Mn-Ga
\cite{Murray-00}. The crystal structure of that martensite (10M) was
defined as a tetragonal one ($a = b \ne c$,\ \ $c/a = 0.94$) with superposed
five-layered modulation. This modulation is known as a kind of 
shuffling of the atoms which has a shape of a static wave propagating along
the [110] direction \cite{Martynov-92,Zayak-03}. This wave has a polarization
vector along [1\=10], i.e., the shuffling is in-plane. The value of 6\% has
been predicted as maximum value for the 10M structure. Recently, new
achievements 
have been reported where 10\% magnetic-field-induced strains were obtained in
Ni-Mn-Ga \cite{Sozinov}. In this case the martensite was found to be the 7M
(14M) 
structure which has an orthorhombic symmetry with superposed 
7-layer modulation. It has even been argued that another martensitic
structure (T) of Ni-Mn-Ga could produce a strain up to 20\%. This specific
structure has 
tetragonal symmetry ($a = b \ne c$,\ \ $c/a \approx 1.25$) without any
modulation. However, this effect has not been observed so far and most
probably the 
non-modulated structure T will not show that kind of MSM behavior. 
The lattice parameters of all these structures are listed in
Table. \ref{tab1}. Each of the three structures, 5M (10M), 7M (14M) and T, can
be the ground state 
depending on the composition of the alloy. Applying stress to
the sample one can induce structural transformations between the 10M, 14M and T
structures \cite{Martynov-92}.

A microscopic explanation of the modulated structures is complicated.
The presence of modulation can be seen from electron diffraction
images where a number of additional spots appear between the main spots defined
by the non-modulated structure. The number of additional spots allows to
define how many atomic planes form a complete period of
modulation. However, a pattern of atomic displacements cannot be defined
uniquely in this way. Whenever some modulated structure is proposed it is a
fit of the measured data to some model. Thus, the 10M structure was
reported to 
have a wave-like pattern of the shuffling and the stability of such structure
was confirmed by recent \textit{ab-initio} calculations
\cite{Martynov-92,Zayak-03}. However, another model formed by long-order 
stacking sequences (instead of the wave-like distortions) can be used (this
time for Ni-Mn-Al which is very similar to Ni-Mn-Ga) to
explain the diffraction patterns with additional spots \cite{Morito-1996}.  
The latter form of the 10M structure has not been calculated so far, but most
probably it will also be stable. Thus the idea here is not to decide which
model is correct and which is wrong, but to presume that the modulation can
be of two kinds: Wave-like or the stacking-like. Which modulation is finally
realized in the sample may also depend on macroscopic strain fields which are
present in the martensitic phase.

In case of the 14M structure the same ambiguity exists concerning the way
how atoms move to form the modulation. Similar to the 10M (5M) structure, the
7-layered martensite also can be presented by the stacking-like modulation
\cite{Morito-1996}. But completely different displacements of atoms
were obtained from recent high-resolution neutron powder diffraction
measurements \cite{Brown}. The modulation in this case is wave-like but with
a very complicated superposition of waves having different periods. Again it
would be productive for further work to assume that both structures are 
to some extent correct.

In this work we report results of \textit{ab-initio} calculations for 
Ni$_2$MnGa in its 7-layer martensite which is formed by the
long-range stacking sequences as it was proposed for the case of Ni-Mn-Al
\cite{Morito-1996}. The wave-like martensite is not considered in this work
but left for future calculations.

\begin{table}[h]
\small
 \caption{\label{tab1} The martensitic structures of Ni-Mn-Ga and
   corresponding chemical compositions collected from  
 Ref. \cite{Koho-2003}. The notations 10M and
 14M stand for the 5-layered and 7-layered modulated structures. Depending on
 how the layers are defined (unit cells or distances between the atomic
 (110)-planes) these structures can be referred to as 5M and
 7M, respectively. In all cases M means modulation. More details and 
 how these martensitic structures can be transformed to each other can  
 be found in Ref. \cite{Martynov-92}.    
}
\vspace{2mm}
   \begin{tabular*}{\textwidth}{@{\extracolsep{\fill}}l c c c }
    \hline
\hline \\
      &  \multicolumn{3}{ c }{Lattice parameters (\AA)} \\
    \hline
 & $\mathrm{5M \; (10M)}$ & $\mathrm{7M \; (14M)}$ &
 $\mathrm{T}$ \\ 
       $a$  & 5.90  & 6.12  &  6.44 \\
    $b$   & 5.90  & 5.78  &  5.52 \\
    $c$  & 5.54  & 5.54  &  5.52 \\
    c/a & 0.94 & - & $\approx$ 1.2 \\
Alloy &  
Ni$_{1.95}$Mn$_{1.23}$Ga$_{0.81}$ & Ni$_{2.02}$Mn$_{1.17}$Ga$_{0.80}$ & 
Ni$_{2.10}$Mn$_{1.06}$Ga$_{0.84}$ \\ 
\hline
   \end{tabular*}
\end{table}

Before discussing details of this work it is necessary to remember
that Ni$_2$MnGa is ferromagnetic system at room temperature ($\mathrm{T_C}
\approx 
380 \; \mathrm{K}$) and undergoes a martensitic phase transformation (MT) at
around 200 K \cite{Webster-84}.  
The initial structure, known as the Heusler structure L2$_1$, has cubic
symmetry. The magnetic moments are mainly localized on the Mn atoms
($\mu_{\mathrm{tot}} = 4.09\; \mu_B$,  $\mu_{\mathrm{Ni}} 
= 0.37\; \mu_B$, $\mu_{\mathrm{Mn}} = 3.36\; \mu_B$, $\mu_{\mathrm{Ga}}
= -0.04\; \mu_B$) \cite{Ayuela-99}. A 
precursor phase transition occurs close to 260 K \cite{Zheludev-96} and
gives rise to a modulated super-structure 3M which is cubic in average but has
some additional shuffling as calculated recently by Ayuela in
Ref. \cite{Zayak-PRB}. The precursor phase transition shows prominent
anomalies in phonon dispersions, magnetization and conductivity
\cite{Zheludev-96,Manosa-01PRB,Khovailo-01JPys}. Below  
the martensitic transformation temperature ($\mathrm{T_M} \approx 202 \;
\mathrm{K}$), Ni$_2$MnGa can be stable in 
different structures, with or without modulation, having tetragonal or
orthorhombic symmetry as listed in Tab. \ref{tab1}. The structural
transformations between the martensitic structures can be
induced by external pressure.

The \textit{ab-initio} total-energy calculations worked well for most of
the structures of Ni$_2$MnGa \cite{Ayuela-99}. The earlier calculations 
allowed to obtain two structures which do not possess any modulation shuffling 
of the atoms. These are 
the initial cubic structure L2$_1$ and a tetragonal martensitic structure
with $c/a \geq 1.2$ \cite{zayak_icomat}. All other structures show a
shuffling of the atoms which cannot be 
explained by taking into account the short-range interactions
alone. Allowing for 
shuffling of the atoms implies that calculations have to be done on relatively
large supercells which is computationally very demanding.
In case of the 10M structure a supercell of 40 atoms had to be
simulated which can still be handled by the density functional method with
sufficient accuracy
\cite{Zayak-03}. The wave-like modulation 10M was calculated and found to be 
related to specific phonon properties of Ni$_2$MnGa as well as
Fermi-surface 
nesting \cite{Zayak-PRB,Veliko-99,Harmon,Claudia}. 
In case of the 7M (14M) structure one would need to do calculations for a
supercell containing 56 atoms. This has been carried out in the present work
in order to define exact structural properties and the criteria of stability
of possible stacking-like 7M (14M) modulations (from now on we use the 14M
notation only).

\section*{2. STRUCTURAL RELAXATIONS}

In order to simulate the 14M structure, we used a supercell, which resembles
the 7-layered structure proposed for Ni-Mn-Al alloys \cite{Morito-1996}. The
supercell consists of seven tetragonal unit cells of 
Ni$_2$MnGa. This allows us to incorporate the full period of the 14M
stacking 
sequences in the supercell. The (110) atomic planes are shifted along the
[1\={1}0] direction so that the modulation propagates along [110].  
Altogether we use 14 atomic planes perpendicular to [110] in order to form the 
supercell. By this construction, two full 7-layered periods
fit into the supercell but also one may choose different variants
of stackings to be simulated. The initial stacking shifts 
were chosen to satisfy the angle $\gamma \approx 3.5^\mathrm{o}$ known for the
monoclinic 14M supercell of Ni-Mn-Al \cite{Morito-1996}. This makes the
shift of two consecutive planes with respect to each other to be 0.3 \AA.

The Vienna {\it Ab-initio} Simulation Package (VASP)
\cite{Kresse-96,Kresse-99} has been used to perform the electronic
structure calculations. The projector-augmented wave formalism (PAW),
implemented in this package,
\cite{PAW,Kresse-99} leads to very accurate results compared to all-electron
methods. Within density-functional theory,
the electronic exchange and correlation are treated by using the generalized
gradient approximation \cite{GGA}; the 3\textit{d} electrons of Ga have
been included as valence electrons. 
The Monkhorst-Pack k-points generation scheme was used with a grid of \ 10
$\times$ 8 $\times$ 2\ points  in the full Brillouin zone of the whole 
supercell.

\begin{figure} 
  \begin{center}  
  \includegraphics*[angle=0,width=16cm]{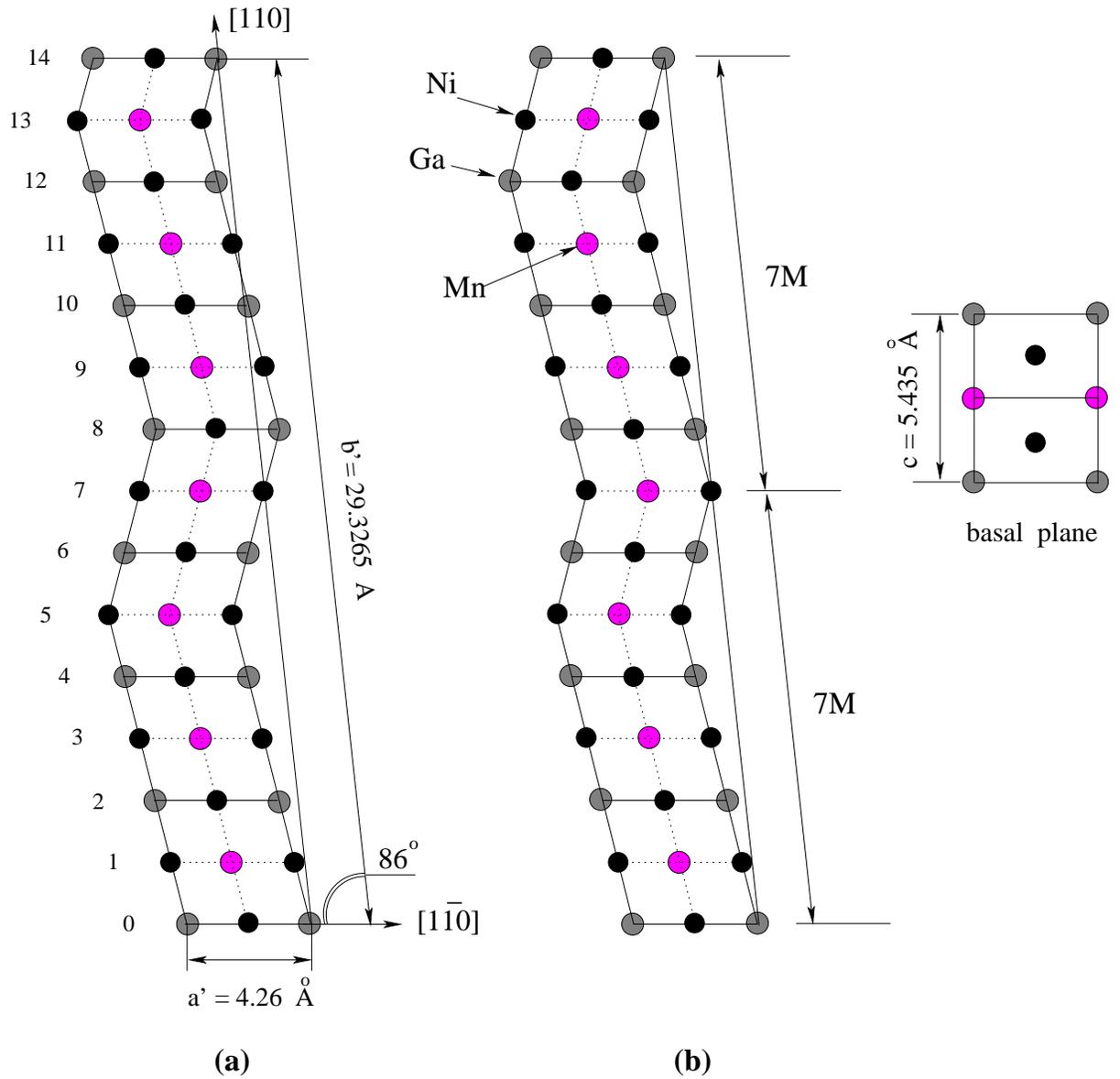}
  \end{center}
  \caption{A.\ Zayak, W. A.\ Adeagbo, P.\ Entel and V. D. Buchelnikov} 
  \label{seven}
\end{figure}

The structural relaxation was performed with respect to the ionic positions,
the shape and the volume of the supercell. We have considered two possible
choices of long-period stacking order sequences: A ``right'' one
($5\overline2$)$_2$ and a ``wrong'' one ($5\overline{3}5\overline{1}$). The
$5\overline{3}5\overline{1}$ structure is different from the normal 14M
structure in the sense that it has an unusual stacking sequence for
modulated structures. We have simulated this specific structure in order to
see whether this modulation can be stabilized. The resulting
14M$_{5\overline{3}5\overline{1}}$ structure is 
shown schematically in Fig. \ref{seven}(a). The relaxation has led to a slightly
different shape and volume but the ``wrong'' initially supplied shape remained
unchanged. We suppose that this means not only the ``wrong'' stacking
corresponds to a local 
energy minimum but that there can be other structures of this kind. The
structure 
shown in Fig. \ref{seven}(a) has a monoclinic symmetry ($a \ne b \ne c$,
$\alpha=\beta=90 \ne \gamma$) with 
the structural parameters: $a_{5\overline{3}5\overline{1}}=4.26\;
\mathrm{\AA}$, 
$b_{5\overline{3}5\overline{1}}=29.32\; \mathrm{\AA}$,
$c_{5\overline{3}5\overline{1}}=5.43\; \mathrm{\AA}$, $\gamma =
86^\mathrm{o}$. The basal planes are shifted with respect to each other by
about 0.34 \AA.

Another structure (14M$_{(5\overline{2})_2}$ ) which has been optimized is
shown 
in 
Fig. \ref{seven}(b). This structure is the one which usually is considered as
the 14M 
stacking-like structure. We note that this structure has also almost remained
unchanged during the relaxation. It appears as if any king of stacking could
be stabilized. However, all possible variants might have different energies
which eventually defines their relative stability. We compare the total
energies of the calculated 
stacking structures with the energies which we already know for other
structures of Ni$_2$MnGa. Figure \ref{ca} shows how the relative total energies
of different structures depend on the $c/a$ ratio which defines the
tetragonality of the structure. The zero energy corresponds to the cubic L2$_1$
structure. One of the curves was calculated for a simple unit cell of
Ni$_2$MnGa with 4 atoms which shows two energy minima corresponding to the
cubic high-temperature L2$_1$ ($c/a=1$) structure and the tetragonal
non-modulated martensite T ($c/a \approx 1.25$). The T structure has a lower
energy compared to the cubic structure as is expected for the zero temperature
calculations. Another curve with just one minimum of energy was calculated
with a supercell of 40 atoms which had a superposed and optimized wave-like
5-layered modulation \cite{Zayak-03}. The minimum at $c/a = 0.94$ corresponds
to the 10M structure which is however metastable with respect to the T
structure. We could not perform $c/a$ calculations for the discussed in this
work stacking-like structures because of the computational costs, but the
energies of this structures are shown in Fig. \ref{ca}. One can see that even
the ``wrong'' stacking leads to a lower energy compared to the cubic
structure. But the ``right'' 14M structure turns out to be more stable. Thus,
we conclude that even though different stacking sequences can be realized,
they differ in total energy.             

Finally we show in Fig. \ref{baca} the total energy surface projected onto the
plane ($b/a$, 
$c/a$), where $a$, $b$ and $c$ are the lattice parameters of the conventional
Heusler structure \cite{Webster-84}. The iso-energy
contour lines show minima corresponding to the cubic  L2$_1$ structure
($b/a=1$, $c/a =1$) and the tetragonal T structure. This energy surface was
calculated for a small unit cell containing 4 atoms and therefore shows no
minima 
corresponding to the modulated structures. But since we are now able to
calculate all modulated structures of Ni$_2$MnGa, it is possible to show the
positions of the minima which would appear if the modulations were taken into
account.

\section*{3. CONCLUSIONS}

We have performed total energy DFT calculations in order to simulate the 14M
structure of Ni$_2$MnGa. Initially we assumed that the modulation of this
structure is stacking-like in contrast to the wave-like one. The latter
modulation was earlier found to coincide with the 10M structure. Here we show
that the 
stacking 
makes the 14M structure stable. However, we found that the stacking sequence
is not unique. Different choices of the stacking order may be made, which
however leads to different total energies. 
The fact that the structure prefers stacking instead of having a
perfect structure agrees with our previous investigations. It has already
been discussed elsewhere that the system 
likes to break the local symmetry which facilitates the formation of hybrid
states of \textit{p}-orbitals of Ga \cite{esomat}. This changes the
distribution of 
the spin-down electronic density of states right at the Fermi level and lowers
the total energy. The role of phonon softening in relation to the modulation
has still to be understood.   
By having the 14M structure simulated we have completed the
list of structures found for Ni$_2$MnGa experimentally (Tab. \ref{tab1}). Now
all these structures have been obtained by DFT calculations. In further work
we will simulate the wave-like 14M structure. Another point to
note is that the relative energies of the simulated structures may be
different in experiment since the martensitic structures of Ni-Mn-Ga are
compositionally dependent. The martensitic domain structure may also
favor some of the structures while all the calculations are done for a single
martensite variant.

\noindent {\normalsize \bf \em Acknowledgments}
\\*[0.5cm]
Financial support by the German Science Council
(Graduate College {\em ``Structure and Dynamics of Heterogeneous
Systems''}) and by the SFB-491 is very much acknowledged. 

{\small
  \bibliographystyle{plain}
  \bibliography{zayak}
}

\newpage \centerline{FIGURE CAPTIONS}

\begin{figure} 
  \begin{center}  
  \includegraphics*[angle=0,width=11cm]{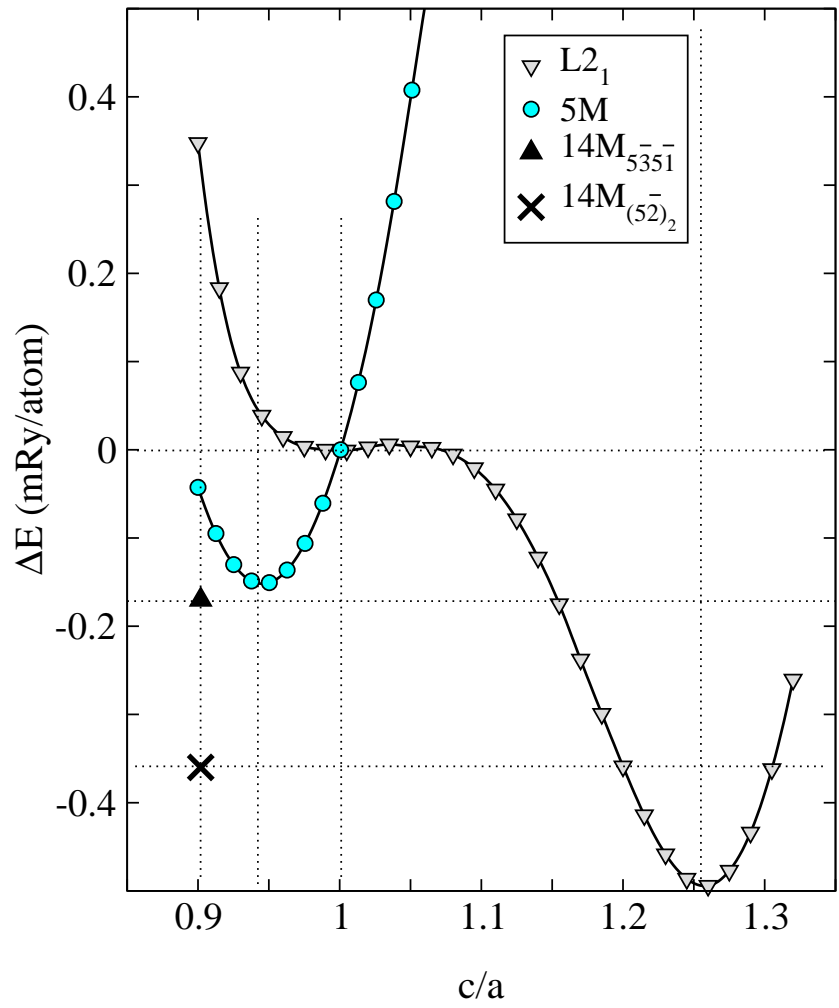}
  \end{center}
  \caption{A.\ Zayak, W. A.\ Adeagbo, P.\ Entel and V. D. Buchelnikov} 
  \label{ca}
\end{figure}

\begin{figure} 
  \begin{center}  
  \includegraphics*[angle=0,width=15cm]{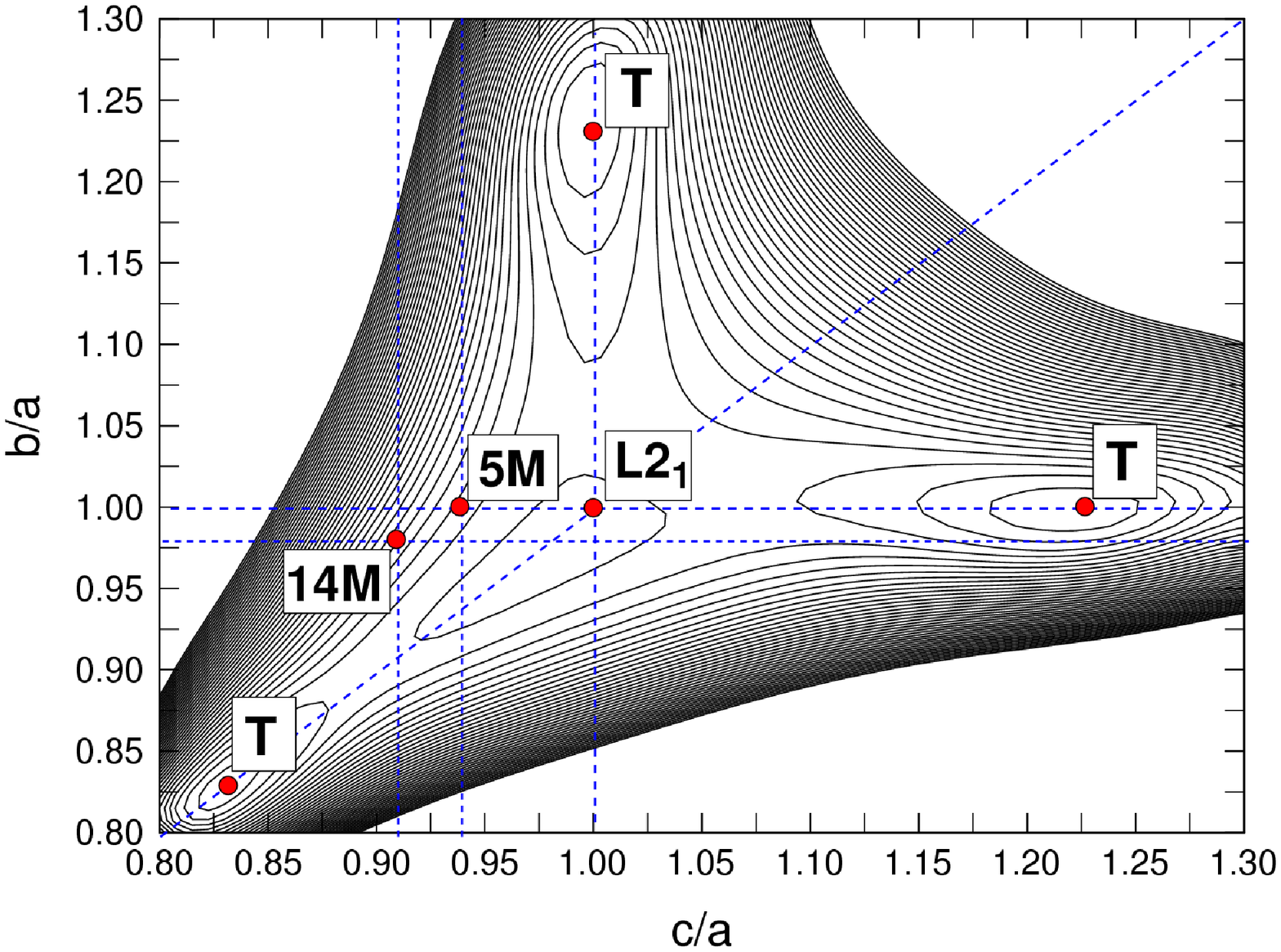}
  \end{center}
  \caption{A.\ Zayak, W. A.\ Adeagbo, P.\ Entel and V. D. Buchelnikov} 
  \label{baca}
\end{figure}

\noindent
Figure 1: \\ 
The modulated stacking-like 14M structures of Ni$_2$MnGa simulated in this
work. The black, gray and white circles show the Ni, Mn and Ga atoms,
respectively. The directions are shown according to the conventional cubic
Heusler structure of Ni$_2$MnGa \cite{Webster-84}. 
(a) The 14M$_{5\overline{3}5\overline{1}}$ structure (the ``wrong'' 14M).
(b) The 14M$_{(5\overline{2})_2}$ structure (the ``right'' 14M).
\\*[0.3cm]
\noindent
Figure 2: \\
The relative total energies of different structures of Ni$_2$MnGa versus the
$c/a$ ratio, which defines the tetragonal distortion of the lattice. The zero
energy corresponds to the energy of the cubic L2$_1$ structure. The 14M
structures are shown here to have lower energies compared to the L2$_1$
structure. The 14M$_{(5\overline{2})_2}$ modulation is more stable then the
14M$_{5\overline{3}5\overline{1}}$ one.
\\*[0.3cm]
\noindent
Figure 3: \\
The total energy surface with iso-energy contour lines plotted to show the
energy minima which can be obtained without taking the modulations into
account. If the modulations are simulated the energy surface would have
additional minima corresponding to the 10M and 14M structures. However it is
impossible to obtain all minima on one ($b/a$,$c/a$) plane because if a
modulation is superposed on the structure, the T structure disappears (see
Fig. \ref{ca}). Also the 10M and 14M modulations have different periods and
characters of shuffling.  
%

\newpage

\end{document}